\documentclass{article}

\usepackage{arxiv}
\usepackage{amsmath}

\usepackage[utf8]{inputenc} 
\usepackage[T1]{fontenc}    
\usepackage{hyperref}       
\usepackage{url}            
\usepackage{booktabs}       
\usepackage{amsfonts}       
\usepackage{nicefrac}       
\usepackage{microtype}      
\usepackage{lipsum}		
\usepackage{graphicx}
\usepackage{natbib}
\usepackage{doi}

\title{Early Warning Signals Appear Long Before Dropping Out: An Idiographic Approach Grounded in Complex Dynamic Systems Theory}
\date{} 					

\author{
\href{https://orcid.org/0000-0001-5881-3109}{\includegraphics[scale=0.06]{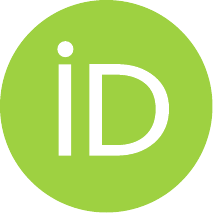}\hspace{1mm}Mohammed Saqr} \\
University of Eastern Finland \\
Joensuu, Finland 
\And
\href{https://orcid.org/0000-0002-9621-1392}{\includegraphics[scale=0.06]{orcid.pdf}\hspace{1mm}Sonsoles López-Pernas} \\
University of Eastern Finland \\
Joensuu, Finland 
\And
\href{https://orcid.org/0000-0003-4039-4342}{\includegraphics[scale=0.06]{orcid.pdf}\hspace{1mm}Santtu Tikka} \\
University of Jyväskylä \\
Jyväskylä, Finland 
\And
\href{https://orcid.org/0000-0001-6040-1939}{\includegraphics[scale=0.06]{orcid.pdf}\hspace{1mm}Markus Wolfgang Hermann Spitzer} \\
Martin-Luther Universität Halle-Wittenberg \\
Halle, Germany 
}

\renewcommand{\headeright}{}
\renewcommand{\undertitle}{Full paper accepted at LAK 2026}
\renewcommand{\shorttitle}{Early Warning Signals Appear Long Before Dropping Out}

\hypersetup{
pdftitle={Early Warning Signals Appear Long Before Dropping Out},
pdfauthor={Saqr et al.}
}

\begin{document}
\maketitle

\begin{abstract}
The ability to sustain engagement and recover from setbacks (i.e., resilience)---is fundamental for learning. When resilience weakens, students are at risk of disengagement and may drop out and miss on opportunities. Therefore, predicting disengagement long before it happens during the window of hope is important. In this article, we test whether early warning signals of resilience loss, grounded in the concept of critical slowing down (CSD) can forecast disengagement before dropping out. CSD has been widely observed across ecological, climate, and neural systems, where it precedes tipping points into catastrophic failure (dropping out in our case). Using 1.67 million practice attempts from 9,401 students who used a digital math learning environment, we computed CSD indicators: autocorrelation, return rate, variance, skewness, kurtosis, and coefficient of variation. We found that 88.2\% of students exhibited CSD signals prior to disengagement, with warnings clustering late in activity and before practice ceased (dropping out). Our results provide the first evidence of CSD in education, suggesting that universal resilience dynamics also govern social systems such as human learning. These findings offer a practical indicator for early detection of vulnerability and supporting learners across different applications and contexts long before critical events happen. Most importantly, CSD indicators arise universally, independent of the mechanisms that generate the data, offering new opportunities for portability across contexts, data types, and learning environments. 

\end{abstract}



\section{Introduction}

Student engagement is a key driver of academic success, with significant consequences for both individuals and society at large \cite{reschly2022jingle,skinner2016engagement, fredricks2019handbook}. On the one hand, engagement acts as a catalyst for well-being, social integration, and academic achievement—often described as the holy grail of learning \cite{reschly2022jingle,Lovelace2017vc, fredricks2019handbook}. As such, engagement has emerged as a central stabilizing factor in leading dropout theories, supported by a substantial body of evidence. On the other hand, disengagement carries serious risks---ranging from lost educational opportunities to poor life outcomes and school dropout \cite{dupere2015stressors, Lovelace2017vc, fraccascia2018resilience, lopezpernas2024dynamics, reschly2022jingle, skinner2016engagement}. Therefore, mapping engagement dynamics and identifying distress signals---early warning of disengagement---within the window of hope enables interventions when they matter most \cite{symonds2024momentary, skinner2016engagement, nazarimehr2020critical}, a gap this study aims to contribute to.

Relatively few studies have explored the longitudinal dynamics of student engagement, particularly the complex, non-linear dynamic  processes that shape how engagement emerges over time \cite{symonds2024momentary,archambault2022student,reschly2022jingle,salmelaaro2021student}. All the more so, our knowledge about disengagement and its dynamics is rather scarce \cite{reschly2022jingle,lopezpernas2024dynamics}. This is a significant gap and an obvious opportunity, as complex dynamic systems' theory offers powerful tools for identifying early signs that precede a transition from engagement to disengagement \cite{fraccascia2018resilience,hilpert2018complex,poquet2023student}. We focus in this study on how the statistical signals of losing engagement and resilience manifest over time when a student is approaching disengagement and ultimately dropping out. In particular, we focus on Early Warning Signs (EWS) and the concept of critical slowing down (CSD). In this context, a tipping point refers to a critical threshold that triggers a sudden and potentially irreversible shift into a disengaged state or a qualitatively new state of performance or dropping out. Prior to reaching this tipping point, the student may exhibit signals of losing resilience (roughly the ability to maintain engagement, i.e., perform daily tasks despite possible adversities) \cite{fredricks2019handbook,helmich2022detecting,nazarimehr2020critical}. Detecting these early indicators makes it possible to intervene long before the transition phase, when efforts to support the student are more likely to succeed and prevent drop-out. This is particularly important given the established value of CSD across several fields, disciplines, and domains of applications where breakthroughs were made into previously intractable problems \cite{dakos2024tipping,milkoreit2018defining,nazarimehr2020critical,olderikkert2016slowing,scheffer2009early}  and received the recognition of the Nobel prize in 2021 \cite{bianconi2023complex}. 

Whereas predicting or forecasting future students’ learning has been performed before ---see \cite{ahmad2022connecting,lu2018applying} for a review--- CSD offers several fundamental advantages over previous approaches to predicting students’ performance. First, CSD is idiographic (single subject) which means it does not aggregate across students but creates a model for each student relying on the student’s own data. In doing so, it offers both specificity and applicability with small sample sizes (even $n=1$)  \cite{dakos2024tipping,milkoreit2018defining,nazarimehr2020critical,olderikkert2016slowing,scheffer2009early,scheffer2010complex}. Second, CSD encompasses universal indicators that have been detected across several contexts “as they occur largely independently of the precise mechanism involved” and that is the “reason for optimism” that they can be far more portable across systems and applicable to more educational domains, contexts and data types \cite{scheffer2009early}. Third, CSD appears and “builds up long before the critical transition occurs”. This temporal advantage of distance (being detected long before) and the predictability of critical moments make EWS unique “temporal predictors” and lay the groundwork for a support mechanism as early and possibly useful as possible \cite{scheffer2009early,scheffer2020critical,scheffer2010complex}. Lastly, if EWS holds true in our case, it is likely that similar methods and assumptions could apply to many other cases in education and disparate systems—just as they have in several other domains—potentially opening the door to a new category of interdisciplinary frameworks and solutions. \cite{dakos2024tipping,milkoreit2018defining,nazarimehr2020critical,olderikkert2016slowing,scheffer2020critical,scheffer2009early}.

Despite established benefits of EWS, and the fact that they were harnessed across a wide array of applications \cite{dakos2024tipping,milkoreit2018defining,nazarimehr2020critical,olderikkert2016slowing,swingedouw2020early}, the worth and utility of such important indicators have yet to be explored in education, a gap which we aim to fill with this study. We examine the value of CSD to detect EWS using a massive dataset of students' math practice in a digital learning environment. In particular, we examine if, when, and how many students' data exhibit EWS such as CSD before they drop out. If learners’ time series consistently show this behavior, it could allow for the development of early detection tools for instability or resilience, and develop early detection methods of a completely different approach to what we already have. As such, our research question (RQ) is:  \textit{Can we detect warning signs of resilience loss in students’ behavior, and if so, when, for how long, and how consistent and accurate are they?}

\section{Theoretical Background}

\subsection{Engagement as a complex system}

Engagement reflects students’ investment, and active participation in learning and includes behavioral, social, and cognitive dimensions \citep{archambault2022student, fredricks2019handbook, skinner2016engagement}. What makes engagement a fertile domain for research and practice is that it is strongly linked to academic achievement, timely graduation and well-being. Engagement ---and in particular the behavioral dimension--- is observable and can be supported to improve students' outcomes, especially when timely interventions are personalized and responsive to individual learning needs \citep{archambault2022student, lopezpernas2024dynamics, reschly2022jingle, skinner2016engagement, symonds2024momentary, tinto2022exploring}. Learning theorists describe engagement with features that are typical of a complex dynamic system where multiple components interact with each other across time \cite{lopezpernas2024dynamics,symonds2024momentary}.  As a process, engagement has a dynamic trajectory that unfolds across time scales and phases, leading to the emergence of patterns \cite{lopezpernas2024dynamics}. Such a dynamic process is often characterized by feedback loops where positive events reinforce future stability (e.g., resilience) and negative adversities may lead to decline of engagement and possibly disengagement \cite{archambault2022student,reschly2022jingle,tinto2022exploring}. 

Disengagement and eventual dropout are rarely abrupt events; rather, they tend to unfold gradually and are often preceded by identifiable distress signals long before a student completely withdraws \cite{dupere2015stressors,zhen2020trajectory}. Despite this, much of the existing research relied on cross-sectional data: compared aggregated averages of engaged versus disengaged others at a single point in time, or traced the aggregate average of a group over time, which obscures the unique trajectories and individual variability that characterize the disengagement process \cite{fredricks2019handbook,lopezpernas2024dynamics,salmelaaro2021student,zhen2020trajectory}. Furthermore, traditional statistical methods are ill-suited for modeling the non-linear dynamics, feedback loops, and emergent behaviors that are central to complex dynamic systems such as student engagement \cite{fraccascia2018resilience,saqr2024why,symonds2024momentary}. These methods also cannot fully account for key features emphasized in engagement theories ---such as distinct phases, cross-scale dynamics, tipping points, and system stability--- limiting their breadth of insights on complex non-linear dynamics \cite{nazarimehr2020critical, symonds2024momentary, lopezpernas2024dynamics}. 

Adopting complex systems theory as a theoretical foundation aligns with a well-established engagement framework and provides a robust lens to examine how engagement unfolds through intricate dynamics across multiple time scales, while also addressing heterogeneity and individual differences \cite{kaplan2020steps,koopmans2020education}. Moreover, endorsing the novel complexity methods offers tools for understanding changes, stability, and resilience, and the possibility to extend our theoretical understanding \cite{dakos2024tipping,scheffer2009early}. In other fields, embracing complexity methods has yielded significant breakthroughs in previously intractable phenomena, such as weather and global warming. Further examples span pharmacology, chemistry, neuroscience, psychology, economy, and ecology to mention a few \cite{dakos2024tipping,fraccascia2018resilience,milkoreit2018defining,nazarimehr2020critical,scheffer2009early}.

\subsection{Temporal Dynamics of Engagement}
From a complex dynamic systems' perspective, engagement can be understood as an emergent state that arises when psychological, environmental, and motivational factors align \cite{symonds2024momentary}. When these conditions converge, they give rise to relatively stable patterns of engagement, known as attractor states \cite{nazarimehr2020critical}. Attractor states represent the system’s tendency to settle into particular modes of functioning—such as sustained engagement or disengagement—that are relatively resistant to changes  \cite{lopezpernas2024dynamics,nazarimehr2020critical}. Attractor states can be consolidated through exposure to favorable conditions and may sustain adverse conditions or face catastrophic failure into disengagement and dropping out under intense setbacks. According to Tinto \cite{tinto2022exploring}, achievement fosters motivation and engagement, which in turn reinforce further motivation; conversely, a lack of achievement can undermine motivation and precipitate disengagement. The end result is self-organization—a process by which the momentary engagement system organizes itself into a state of stability i.e., sustained resilient engagement \cite{tinto2022exploring, symonds2024momentary, lopezpernas2024dynamics}.

Engagement—and the organizational patterns that underpin it unfolds across multiple time scales: moments, days, and entire programs \cite{salmelaaro2021student}. These time scales are deeply interconnected and influence one another to form the fabric of the engagement process \cite{nelson2006dynamics,scheffer2009early,scheffer2020critical,scheffer2010complex}. Momentary experiences accumulate over time to contribute to the emergence of stable patterns of engagement (parts-to-whole dynamic). At the same time, stable patterns—or attractors—shape and constrain moment-to-moment experiences (whole-to-parts dynamic), reinforcing the system’s tendency to maintain engagement and resist changes \cite{nelson2006dynamics,symonds2024momentary}. However, responses to change within complex systems are often non-linear. Shifts at shorter time scales can lead to unexpected, drastic emergent behaviors over time, and these changes vary by each individual \cite{nazarimehr2020critical,scheffer2009early,scheffer2010complex,scheffer2020critical}. Heterogeneity is a fundamental characteristic of human behavior and, as such, it is entirely expected that students vary widely in their stability: some remain deeply and consistently engaged even under pressure, demonstrating resilience and adaptability, while others are far more susceptible to disruption, where even minor setbacks can trigger disproportionate responses or cascading disengagement \cite{bryan2021behavioural,lopezpernas2024dynamics}.

\begin{figure*}[!ht]
\centering
\includegraphics[width=.8\linewidth]{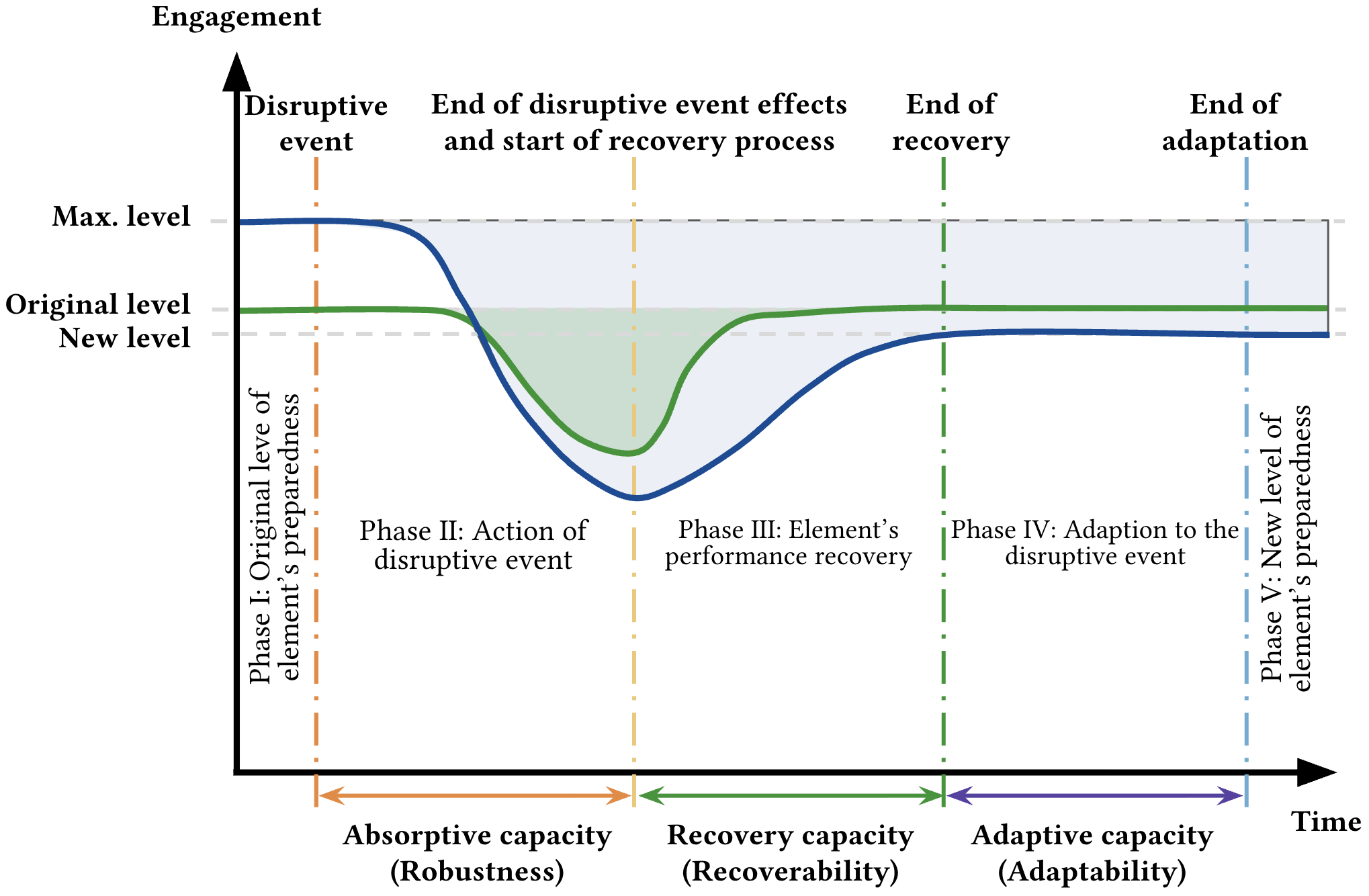}
\caption{The figure highlights phases of resilience: initial state, the disruption, recovery, adaptation, and the establishment of a new engagement level. Resilience—through absorptive, recovery, and adaptive capacities—shapes how quickly and effectively engagement recovers. A resilient student (green) recovers faster and reaches a higher engagement level, while a less resilient student (blue) faces a longer decline, slower recovery and a lower final level \citep{rehak2018resilience}}.
\label{fig:CSD_fig1}
\end{figure*}

The capacity to maintain function —stay engaged— is best understood through the lens of resilience—the system’s (students in our case) ability to absorb setbacks, maintain core functions (sustain engagement), and recover without transitioning into a less functional regime (get disengaged and drop out) \cite{fraccascia2018resilience,skinner2016engagement,thurner2018introduction}. As resilience weakens, the system’s restoring forces diminish, and the setbacks produce longer, more volatile departures from the attractor engaged state. Put another way, when a student loses resilience, they struggle to recover from disruptions that would have previously been manageable \cite{fraccascia2018resilience,nazarimehr2020critical}. As illustrated in Fig.~\ref{fig:CSD_fig1}, a resilient student (represented as the green line) is subjected to a setback (e.g., a poor exam performance or a challenging subject), and we see a noticeable decline in their engagement. Over time, the student begins to recover, with engagement gradually improving as they adjust to the challenge due to their strong resilience capacity. Eventually, the student adapts and returns to the same level of engagement. The less resilient student (blue) experiences a deeper decline in engagement when faced with a challenging situation at school. Due to the student’s weak resilience and limited adaptive capacity, the student recovery is slower and less complete leading to a transition toward sustained disengagement.

\subsection{Detecting Transitions in complex systems with CSD indicators}

\begin{figure*}[!ht]
\centering
\includegraphics[width=.8\linewidth]{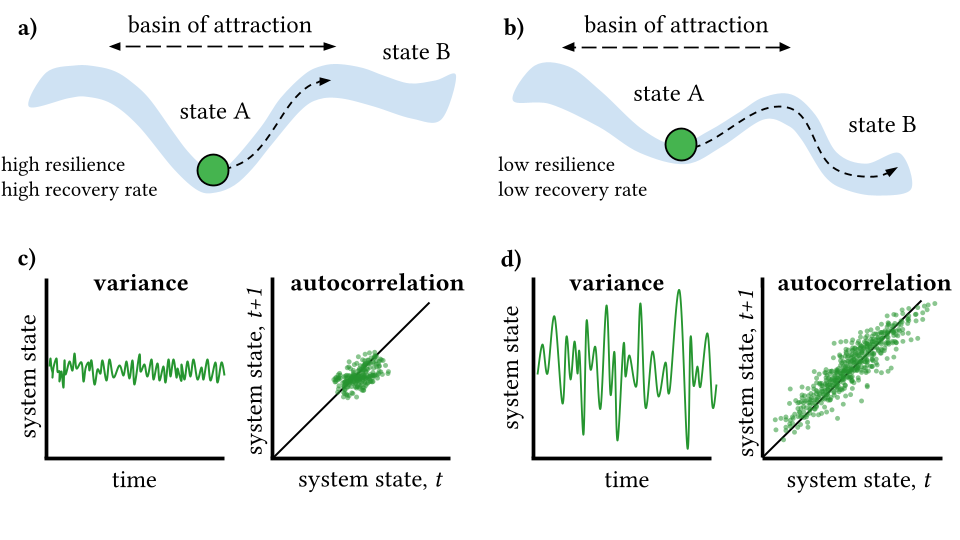}

\caption{Panels (a) and (b) illustrate the system’s stability landscape. In panel (a), the system resides in a stable state (state A), characterized by high resilience to perturbations and a fast recovery rate; disturbances do not push the system toward the alternative state (state B) because of the deep basin of attraction. In panel (b), the system approaches a critical transition: the basin of attraction becomes shallower and narrower, increasing the likelihood that perturbations will shift the system to an alternative stable state (state B). Panels (c) and (d) show the corresponding time-series behavior. In panel (c), a stable system exhibits low variance and low autocorrelation, indicating small, rapidly damped fluctuations. In panel (d), as the system nears a tipping point, both variance and autocorrelation increase, reflecting larger and more persistent fluctuations. This pattern signals critical slowing down (CSD) and serves as an early warning of an impending transition \cite{scheffer2009early}.}
\label{fig:CSD_fig2}
\end{figure*}

Critical transitions in complex systems are typically driven by bifurcations — \textit{points} where slow, continuous losses in resilience cause a sudden collapse (i.e., loss or failure of the ability to sustain functioning) and a shift from an engaged to a disengaged state, reflecting a fundamental change in the student engagement behavior \citep{helmich2022detecting, nazarimehr2020critical, scheffer2010complex}. As a student nears such a tipping point, CSD occurs: the return rate to a stable state slows, disturbances (disengagement) persist longer (which can be observed as rising lag-1 autocorrelation), and weakened restoring forces allow random perturbations to accumulate (reflected as increasing variance) \citep{dakos2024tipping, helmich2022detecting, milkoreit2018defining, nazarimehr2020critical, olderikkert2016slowing}. In the frequency domain, this is observed as spectral “reddening” where power shifts toward lower frequencies as slower fluctuations dominate \citep{nazarimehr2020critical}. These same dynamics can also alter skewness, reflecting emerging asymmetry in the system’s potential, while flickering between alternative states under strong perturbations increases kurtosis \citep{dakos2024tipping, helmich2022detecting,  nazarimehr2020critical}. The process is further demonstrated in Fig.~\ref{fig:CSD_fig2}. These changes offer an opportunity. Detecting a system’s vulnerability to losing resilience—through early signals such as slower recovery from disruptions or reduced responsiveness—offers an opportunity to anticipate and intervene before a full transition into disengagement occurs \citep{dakos2024tipping, helmich2022detecting, milkoreit2018defining, nazarimehr2020critical, olderikkert2016slowing, scheffer2009early}.

Figure \ref{fig:CSD_fig2} illustrates these dynamics using the concept of a basin of attraction—a region in the system’s state space where it tends to remain or return to after setbacks. The stable attractor states represent typical behavioral patterns, such as sustained engagement (state A) or disengagement (state B). When resilience is high, the basin of attraction is deep and wide, which keeps the student firmly within the engaged attractor despite setbacks. As resilience declines and the student approaches a bifurcation, the basin becomes shallower and narrower, weakening the restoring forces that maintain the current state. This leads to slower recovery, increased variance, and higher autocorrelation in fluctuations, signaling a CSD and an increased risk of transitioning to a less functional state, i.e., sustained disengagement.

One of the most intriguing findings about complex systems is the universality of some of their signature behaviors and the observation that diverse systems exhibit similar patterns near their critical points \cite{milkoreit2018defining,scheffer2009early}. Be it an ecological system facing collapse, a financial market on the brink of a crash, or a brain approaching a seizure or falling into a depressive episode, these seemingly unrelated systems display strikingly similar behaviors—such as power laws or scaling phenomena—as they near critical transitions, where their inner workings often become less important than their collective dynamics \cite{dakos2024tipping,milkoreit2018defining,nazarimehr2020critical}. This cross-domain consistency suggests that CSD reflects a fundamental principle of how systems behave near critical thresholds. It stands to reason that ---by extension---  the same framework may illuminate engagement dynamics where critical events such as disengagement could also be preceded by detectable early warning patterns.

In this study, we will examine the detection of these CSD indicators to capture the decline in resilience before a critical transition occurs. The ability to detect approaching bifurcations in student engagement represents a significant opportunity to shift from reactive to proactive intervention strategies. Such a method would enable timely, targeted interventions designed to improve student resilience and prevent the collapse into academic disengagement, shifting support from a reactive to a proactive model. The increasing availability of high-frequency digital learning data (clickstream data, learning management system interactions, physiological measures) provides unprecedented opportunities to monitor the statistical properties that characterize approaching critical transitions.

\subsection{Predicting Student Dropout}
Since 2011, learning analytics has extensively used data from digital learning environments to predict struggling students to enable timely interventions when it still matters \citep{ahmad2022connecting, lu2018applying}. While numerous studies achieve strong performance predictions within individual courses, replicating these predictions using similar data and methods remains challenging due to contextual variability, learning design, and individual differences \citep{ahmad2022connecting, lu2018applying}. Attempts to improve model portability with complex data and modern algorithms have had limited success across contexts \citep{conijn2016predicting, saqr2024idiographic}. These difficulties in replication may stem from existing models relying on aggregate data, which obscures individual differences, leading to good prediction performance in aggregate but poor performance for specific learners or new contexts \citep{saqr2024idiographic}. Another problem is that differing contextual environments make replication far more difficult than once hoped \citep{ahmad2022connecting, lu2018applying}. 

CSD offers several advantages over previous predictive models. First and foremost, CSD relies on single-subject models and requires no more than a single student's time-series data with sufficient depth and resolution to model. This brings several opportunities for modeling single students' data e.g., physiological data, behavioral temporal data, interaction data, or passive sensor data from, for instance, attention or screen-time data \citep{dakos2024tipping, milkoreit2018defining, scheffer2009early}. Moreover, CSD encompasses generic universal indicators that are context-independent and can manifest across many systems. This, of course, does not make CSD inherent to every system but subject to empirical verification. Third, CSD signals emerge long before and far earlier than those detected by traditional methods, providing ample time for intervention \citep{dakos2024tipping, milkoreit2018defining, scheffer2009early}. 
Fourth, traditional methods for identifying dropout risk often rely on detecting changes in absolute levels of behavior or reaching predefined thresholds (e.g., attendance, grades). However, such indicators may signal problems only after a transition into disengagement is already underway—by which point effective intervention may be too late. In contrast, CSD  indicators focus on detecting subtle changes in the system’s dynamic properties—such as slowing recovery, rising variance, and autocorrelation—before observable shifts in engagement occur. This allows for earlier identification of vulnerability and creates a window for preventive action prior to full disengagement or dropout.
Last, the minimal data requirements for modeling make it well-suited to the growing landscape of multimodal physiological data, including heart rate variability, electrodermal activity, pupil dilation, or passive sensor data such as screen time, keyboard dynamics, or gaze tracking.

\subsection{Critical slowing down in the literature}

Over the past two decades, CSD has been successfully applied to anticipate critical transitions across a wide variety of domains \cite{scheffer2009early, nazarimehr2020critical}. In ecology, CSD indicators have signaled ecosystem collapse such as lake eutrophication or desertification \citep{Dakos2015-zx}. In climate science, they have been used to detect early signs of large-scale shifts in ocean circulation and ice-sheet stability \citep{milkoreit2018defining}. In neuroscience and psychology, CSD has been observed in the onset of epileptic seizures, depressive episodes, and other critical transitions in mental health \cite{helmich2022detecting}. Similar methods have also been applied to financial markets, where high variability preceded large market crashes  \cite{dakos2024tipping}. Across these domains, CSD has consistently provided advance warning of resilience loss leading up to critical transitions, often long before traditional forecasting methods could do so. These findings underscore the promise of applying CSD in education, where vulnerability in resilience may likewise mark a tipping point in a learner’s data.

Despite the demonstrated value of CSD across many scientific fields, its potential to detect resilience loss as a tipping point of vulnerable (dis)engagement in education has not yet been explored. In this study, we leverage large-scale time series data from students’ interactions in a digital learning environment for mathematics to examine whether CSD indicators emerge before dropout. Specifically, our aims are to  (i) investigate how and when EWS appear across students using CSD indicators before dropping out happens, 
(ii) determine how early these signals can be detected relative to dropout events, (iii) assess the cases in which false positive detections occur. 


\section{Method}
\subsection{Context}

The context of this study is a digital learning environment known as \textit{bettermarks} for mathematics, used in Germany and the Netherlands \citep{spitzer2025basic}. The digital learning environment targets students from secondary education (ages 9--18) across both academic and vocational tracks, making it one of the most widely used digital mathematics tools in German-speaking classrooms (annually used by $\approx$ 300,000 students). The system is structured into more than 100 digital textbooks aligned with national curricula. These textbooks cover the full spectrum of school mathematics---from arithmetic and fractions to algebra, functions, geometry, and statistics. Each textbook is divided into chapters and practice sets, with individual problem sets typically containing 8--9 problems that require one to three solution steps each. Across the platform, this amounts to more than 100,000 interactive problem steps.

One feature of \textit{bettermarks} problems is their dynamic generation: when students repeat a problem set, the system re-parameterizes items (e.g., by varying numbers or contexts), ensuring that practice remains meaningful rather than encouraging memorization. Problems within a set must be completed sequentially, reinforcing structured problem solving (for other features see \citep{spitzer2025basic}).
Students receive immediate step-level feedback (correct/incorrect) and can retry each step once, providing the opportunity to learn from errors without unlimited guessing. Performance is summarized through an error rate, calculated as the ratio of actual errors to the maximum possible errors in a set. This metric accounts for retries and enables both teachers and researchers to track performance at a granular level.
The platform is designed for teacher-led instruction. Teachers and students are both registered, allowing teachers to assign problem sets for classroom work or homework. Teachers receive within-system performance summaries (e.g., error rates per problem set) and can monitor student progress in real time. Students, in turn, benefit from step-level feedback, set-level performance reports, and access to repeated practice opportunities. This dual perspective supports formative assessment and data-informed teaching practices.


\subsection{Early Warning Signal (EWS) Detection}

The aim of this study is to capture the emergence of early signals of disengagement and, more importantly, the catastrophic form of it, dropping out of math practice and homework using EWS \cite{dupere2015stressors,skinner2016engagement,zhen2020trajectory}. Put another way, our aim is to detect signs that a student is going to drop out long before that happens using the statistical properties (CSD indicators) of their math practice data to detect EWS.

To detect EWS of potential critical transitions (i.e., dropout), a suite of statistical indicators was computed using the R package \texttt{codyna} \citep{Tikka2026-gu} within an expanding window of size 50 math practice problems (note that a large window is usually required, with 50 being the most common) \cite{dakos2024tipping,nazarimehr2020critical,EWSmethods}. Each metric captures a distinct statistical EWS feature associated with CSD near a tipping point. All metrics were computed using an expanding window where data points are added sequentially to an accumulating history, standardizing each new assessment against the running mean and standard deviation of the previous window. Trends in these indicators were then examined to assess whether the system exhibited signs of CSD  prior to transition points\cite{dakos2024tipping,milkoreit2018defining,nazarimehr2020critical,scheffer2009early}. Autoregressive coefficient of order 1 ($AR(1)$) quantifies the correlation between a data point and its immediate predecessor. $AR(1)$ reflects the system's short-term memory. $AR(1)$ was estimated by fitting an autoregressive model of order 1 using ordinary least squares to the demeaned data within each window. An increase in $AR(1)$ toward 1 indicates slower recovery from perturbations, consistent with CSD and a critical transition. Standard deviation ($SD$) measures the dispersion of data points around the mean. As a system approaches a tipping point, fluctuations tend to increase, resulting in higher $SD$. This rise in variability serves as an indicator of reduced system stability. Skewness characterizes the asymmetry of the distribution of values. Deviations from symmetry may reflect directional shifts in the system's dynamics, signaling a warning of a shift. Kurtosis describes the "tailedness" of the distribution, capturing the frequency and extremity of outlier values. An increase in kurtosis indicates a greater likelihood of extreme deviations, often associated with heightened instability near a transition. Coefficient of variation ($CV$) is calculated as the standard deviation divided by the mean ($CV = SD / \mu$). $CV$ provides a normalized measure of variability, allowing comparison across windows with different scales. An increasing $CV$ suggests growing relative variability and declining resilience. Return rate ($RR$) is defined as \( RR = 1 - AR(1) \), and quantifies the speed at which the system returns to equilibrium following a disturbance. A decreasing return rate reflects slower recovery and is considered a direct signal of CSD \cite{dakos2024tipping,milkoreit2018defining,nazarimehr2020critical,scheffer2009early}.

Given that assessment items vary in difficulty, and not all contribute equally to measuring ability, a relative mean score (\textit{rel\_score}) was used to better capture student performance and to account for varying item difficulty based on psychometric item response theory. The variable  \textit{rel\_score} subtracts the mean score of all students for a given item from an individual student's score, centering performance around 0 for average, positive for above-average, and negative for below-average. In doing so, we prioritize relative proficiency over raw score, which aligns with psychometric practices that consider item difficulty for interpreting test scores, especially in diverse assessment and varying difficulty contexts. Formally, for a given student $i$ and item $j$, the relative score was calculated as:

\[
\mathrm{rel\_score}_{ij} = x_{ij} - \bar{x}_{\cdot j},
\]
where $x_{ij}$ is the score of respondent $i$ on item $j$, and $\bar{x}_{\cdot j}$ is the mean score for item $j$ across all respondents.

To characterize the temporal distribution of the detected EWS for each student, two metrics were computed: the \textit{Signal Spread} and the \textit{Signal Density}. The Signal Spread was calculated to measure the total duration over which signals occurred relative to the student's entire activity period. A low Signal Spread indicates that all detected signals were tightly clustered in a narrow portion of the student's activity. To quantify the intensity of signals within this span, the Signal Density was calculated, which measures the proportion of time points within the warning window that contained an active signal. A high Signal Density (approaching 1.0) signifies a highly concentrated and persistent burst of warnings.

Cases where EWS were detected exclusively within the first half of each student's dataset were isolated for further analysis. For these selected cases, the time series was partitioned into distinct \textit{before} and \textit{after} periods relative to the identified warning point.
To quantify the change, we computed the relative percentage change in the mean between these two periods using the formula:
$$\text{Change} \% = \frac{(\text{Mean}_{\text{after}} - \text{Mean}_{\text{before}})}{|\text{Mean}_{\text{before}}|} \times 100$$
Furthermore, while assessment of formal statistical significance is not a conventional requirement in regime shift analysis, we computed t-tests as an exploratory measure to further quantify the magnitude of the observed changes. A Welch's two-sample t-test was performed to determine if the observed change in means was statistically significant ($p < 0.05$).

While the core EWS analysis was executed on an individual student basis (using idiographic analysis), the findings reported here represent example cases and the aggregated statistics compiled from all datasets, given the impracticability of reporting on each of the 9,401 distinct students  individually.

\section{Results}

\subsection{Descriptive statistics}

Our dataset focused on dropouts and contained 1,673,541 observations from 9,401 students representing students who dropped out early before completing their full math practice. The primary variable of interest, $\text{rel\_score}$, had a mean of $-0.02$ (SD = 0.30), median = $-0.08$, and ranged from $-0.94$ to 0.98 across the entire dataset of all students. The distribution was slightly positively skewed (skewness = 0.59) and relatively flat (kurtosis = $-0.18$), indicating that most attempts had scores clustered just below the center—consistent with the centering of individual responses. Time in seconds had a mean of 369.84 (SD = 4,285.70), with an extreme range from $-$1,208,242 to 4,421,593, suggesting possible anomalies in how the system recorded usage time; due to this, the variable was excluded from further analysis. On average, each student solved 178.02 questions (SD = 40.70). Students were active for an average of 6.94 days per month (SD = 1.89) and solved 4.69 questions per active day (SD = 2.71). The mean relative score per student (i.e., the average of each student’s  $\text{rel\_score}$) was $-0.02$ (SD = 0.11), with a median of $-0.02$ and a range from $-0.36$ to 0.47. The distribution was slightly positively skewed (skewness = 0.23 and kurtosis = 0.19).

\subsection{Overall detection}
\begin{figure*}[ht]
\centering
\includegraphics[width=0.85\linewidth]{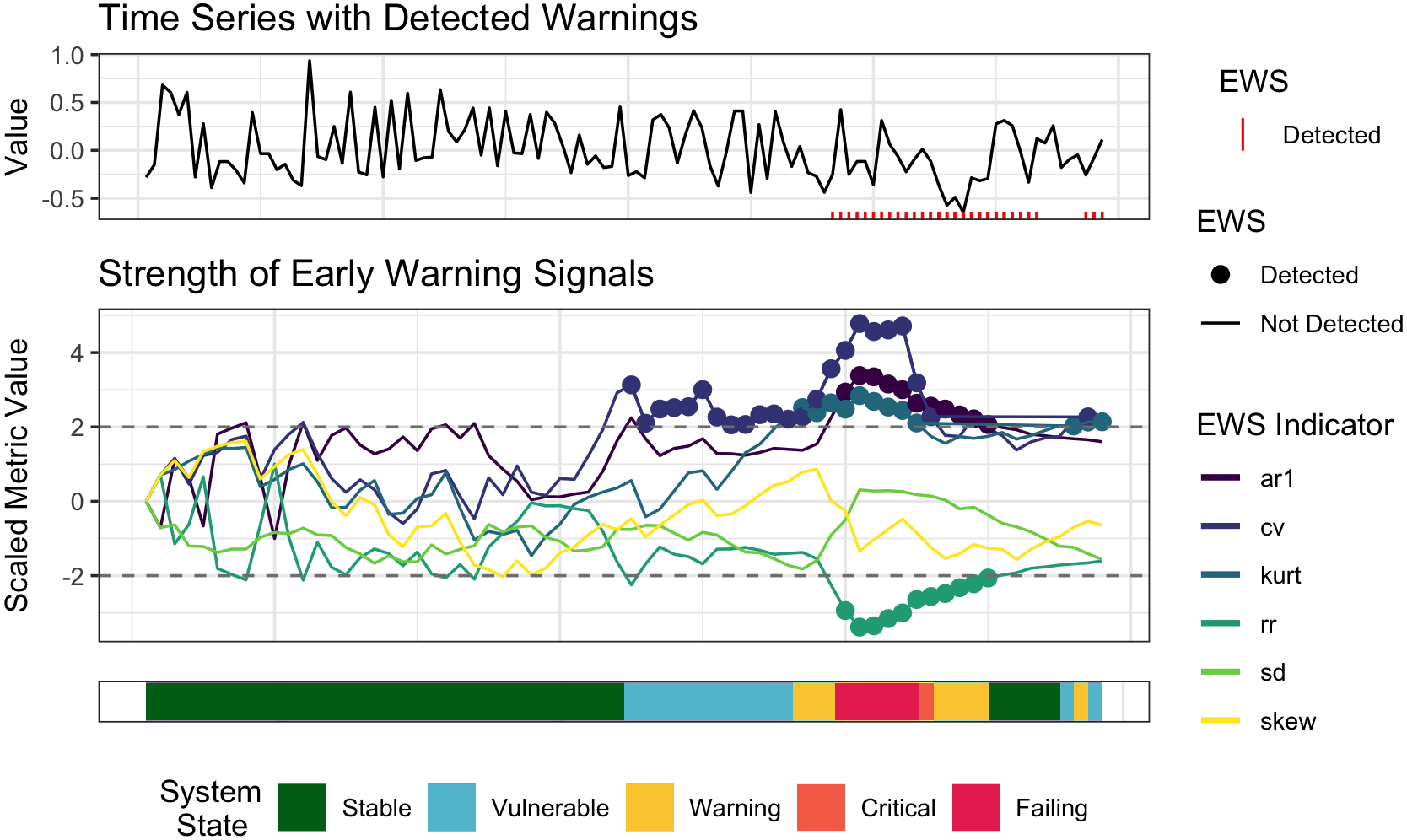}
\caption{A time series with detected warnings (top panel), the scaled metric values of various early warning indicators over time (middle panel), and the corresponding system state inferred from these signals (bottom panel). The top panel presents a raw time series, with red dashed lines indicating periods where EWS were detected. The middle panel plots the scaled values of several EWS indicators (\textit{ar1}, \textit{cv}, \textit{kurt}, \textit{rr}, \textit{sd}, \textit{skew}) over time. Darker dots on these indicator lines signify detected EWS. The bottom panel visually represents the system State transitioning from Stable to Vulnerable, Warning, Critical, and Failing, based on the combined EWS signals.
}
\label{fig:CSD_fig3}
\end{figure*}

We first start by showing a typical student in Fig. \ref{fig:CSD_fig3} as an example.
While the student had some fluctuation in performance during the initial practice, their relative score remained fairly stable. At problem 85 (prop 0.72), a pattern of volatility (coefficient of variation) appeared, suggesting that the student was beginning to struggle. Soon after, additional warning signals followed. The system’s memory lengthened as both autocorrelation ($AR(1)$) and recurrence rate ($RR$) increased, indicating that the student was no longer recovering from mistakes as readily as before. Simultaneously, the distribution of scores began to shift, with rising kurtosis reflecting increasingly extreme, erratic fluctuations in performance. The combined evidence from volatility, slowed recovery, and distributional changes all pointed to the fact that the student’s stability was collapsing. Soon after, the student dropped out, shown as discontinuation of practice in the top panel.

Out of 9,401 students analyzed, 8,288 met the detection criteria (at least 2 successive points where the standardized metric is greater than 2 standard deviations (less than -2 for return rate)) with an average of 23.72 detections per dataset (detection rate 88.2\%). Single warning points were disregarded to reduce random noise, avoid false positives and ensure that only sustained and potentially meaningful changes were captured. Of the 222,986 total warnings (mean = 23.72, median = 16, range 0--292), 173,025 (77.7\%) occurred in the second half, while only 49,638 (22.3\%) appeared in the first half with 14.5\% showing signals confined exclusively to the first half. This skew toward later detections is a signature of critical slowing down, where systems accumulate warning signs as they approach a collapse or transition. The quarterly breakdown reinforced this late-stage concentration. The first quarter accounted for just 134 warnings—0.1\% of the total—while the second quarter contributed 49,504 (22.2\%). Activity peaked in the third quarter with 88,519 warnings (39.7\%) and remained high in the fourth quarter with 84,506 (37.9\%). Fig. \ref{fig:CSD_fig3} presents an example for one single student for whom EWS were detected.


Timing metrics aligned with this pattern. The first detection occurred, on average, at 0.47 of the series length (median 0.44), close to the midpoint. However, the strongest signals consistently emerged later, beginning on average at 0.56 (median 0.537), peaking around 0.643 (median 0.634), and tapering off at 0.71 (median 0.711). This consistent trend illustrates that signals intensify and concentrate toward the end of the series, with earlier detections representing exceptions rather than the norm.

\begin{table}[!ht]
\centering
\caption{Detailed Timing (Detected Datasets)}
\small
\begin{tabular}{l r r}
\toprule
\textbf{Measure} & \textbf{Mean} & \textbf{Median} \\
\midrule
First Detection (prop)        & 0.470 & 0.439 \\
Strongest Signal Start (prop) & 0.559 & 0.537 \\
Strongest Signal Median (prop)& 0.643 & 0.634 \\
Strongest Signal End (prop)   & 0.710 & 0.711 \\
Signal Density          & 0.647 & 0.600 \\
Mean Signal Spread            & 0.144 & 0.129 \\
\bottomrule
\end{tabular}
\label{tab:detailed-timing}
\end{table}

On average, each dataset was flagged by 2.76 metrics (SD = 1.56), with a median of 3.00, indicating that the typical case involved signals from three distinct metrics and a mean of 23.72 detection points. This reflects a consistent consensus among indicators of resilience loss, where multiple dimensions of slowing down converged to mark weakening system stability. The performance of individual metrics further illustrates how different indicators capture resilience loss in complementary ways: slowing recovery ($ar1$, $rr$, $sd$), nonlinear distortions ($skew$, $kurt$) or volatility ($cv$). Autocorrelation ($ar1$) and recurrence rate ($rr$) were the most reliable indicators, each detecting 51.7\% of datasets. Their role lies in directly capturing the slowed recovery characteristic of critical slowing down: as systems lose resilience, disturbances linger, and these metrics detect the persistence of state deviations. Standard deviation ($sd$), with a nearly identical detection rate of 50.5\%, complements this by quantifying how weakened restoring forces allow fluctuations to accumulate, thereby marking an erosion of stability.

\begin{table}[!ht]
\centering
\caption{Metric Summary}
\small
\begin{tabular}{l r r}
\toprule
\textbf{Metric} & \textbf{Detection rate (\%)} & \textbf{Mean count} \\
\midrule
Autoregressive Coefficient AR(1)   & 51.7 & 7.44 \\
Return Rate (RR)    & 51.7 & 7.44 \\
Standard Deviation (SD)    & 50.5 & 8.27 \\
Kurtosis  & 46.6 & 9.83 \\
Skewness  & 40.2 & 7.84 \\
Coefficient of variation (CV)    & 35.2 & 11.67 \\
\bottomrule
\end{tabular}
\label{tab:metric-summary}
\end{table}

Higher-moment measures provided a different lens on resilience loss. Skewness 
detected 40.2\% of datasets, signaling the emergence of asymmetry in system dynamics as the effective potential begins to tilt under weakening stability. Kurtosis
, which identified 46.6\%, pointed to the presence of flickering and heavy-tailed fluctuations, consistent with systems intermittently crossing between alternative states as resilience decays. The coefficient of variation ($CV$) was the least prevalent at 35.2\%, yet when triggered it produced the highest mean volume of warnings (11.67 per dataset). This suggests that $CV$ acts as a highly sensitive indicator of volatility once resilience has degraded sufficiently, amplifying its signal in the final approach to collapse. By contrast, the more reliable detectors—$AR(1)$ and $RR$—produced fewer warnings on average (7.44), reflecting their steadier role in monitoring the gradual persistence of perturbations.

The temporal dynamics of warning signals reveal a consistent tendency for clustering in narrow bursts rather than diffuse distributions. Signal Spread---the extent to which warnings are dispersed across a series--- was generally low (mean = 0.14, SD = 0.11; MAD = 0.12), with a positively skewed distribution (Skewness = 0.95). This indicates that most systems exhibited tightly bounded warning activity, while only a small minority showed broader dispersal. Such low spread is consistent with the localized buildup expected under critical slowing down.
Signal density was comparatively high (mean = 0.66, SD = 0.22) with values spanning 0.19 to 1 and a nearly symmetric distribution (Skewness = 0.29) which indicates high concentration of warning signals. On average, two-thirds of all warnings in a series occurred within a single focused window, underscoring that signals typically accumulate abruptly rather than gradually. Together, these findings suggest that resilience loss manifests in localized bursts of warnings that intensify as systems near their critical thresholds.

\subsection{Warning not-associated with dropping-out}

Approximately 14.5\% of all students exhibited EWS confined to the first half of the dataset, without subsequently dropping out. These cases raise an important question: were these signals false positives, random noise, or indicators of other significant changes? Upon further analysis, it was found that many of these cases did in fact precede major changes, suggesting that the EWS were not spurious (Table \ref{tab:agg}). Rather than noise, these signals often marked the onset of transitions. Specifically, 58.9\% of these students experienced an increase in their math practice scores following the warning signal, while 41.1\% saw a decline.

\begin{table*}[!ht]
\centering

\caption{Aggregated Change Summary by Size, Direction, and Significance}
\label{tab:agg}
\small
\begin{tabular}{llcccc}
\toprule
\textbf{Change Size} & \textbf{Direction} & \textbf{Total Cases} & \textbf{Significant (p$<$0.05)} & \textbf{Marginally Sig. (p$<$0.1}) & \textbf{Percent of Total} \\
\toprule
\textbf{Small}   & Decrease & 118 & 0 (0.0\%) & 1 (0.8\%) & 11.6\% \\
        & Increase & 109 & 4 (3.7\%) & 4 (3.7\%) & 10.7\% \\
        \midrule
\textbf{Medium}  & Decrease & 147 & 27 (18.4\%) & 37 (25.2\%) & 14.4\% \\
        & Increase & 256 & 70 (27.3\%) & 107 (41.8\%) & 25.1\% \\
            \midrule
\textbf{Large}   & Decrease & 131 & 51 (38.9\%) & 63 (48.1\%) & 12.8\% \\
        & Increase & 207 & 121 (58.5\%) & 141 (68.1\%) & 20.3\% \\
            \midrule
\textbf{Massive} & Decrease & 24  & 11 (45.8\%) & 16 (66.7\%) & 2.3\% \\
        & Increase & 29  & 14 (48.3\%) & 17 (58.6\%) & 2.8\% \\
\toprule
\textbf{Total}   & All      & 1021 & 298 (29.2\%) & 386 (37.8\%) & 100\% \\
\bottomrule
\end{tabular}
\\ 
\footnotesize \textit{Note.} Change \% categories are defined based on the absolute percentage change $|x|$ as follows: Small: $|x| < 25\%$; Medium: $25\% \leq |x| < 100\%$; Large: $100\% \leq |x| < 1000\%$; Massive: $|x| \geq 1000\%$.

\end{table*}

In terms of magnitude, the majority of transitions involved medium to large changes: 39.5\% experienced a change of 25\% to 100\% in their scores, 33.1\% saw large changes between 100\% and 1000\%, and 5.2\% had increases exceeding 1000\%. Overall, 72.6\% of these transitions involved at least a medium-sized change, supporting the interpretation that early warning signals frequently preceded substantive shifts in student behavior. While definitions of regime change may vary---depending on whether one sets the threshold at a 25\% or 100\% change---the key conclusion remains: many of these early warning signals corresponded to genuine precursors of major shifts in math practice scores. Importantly, these changes are consistent with the notion that EWS capture instability of the current equilibrium, without necessarily indicating the direction of the impending transition. Statistical significance tests reinforced this interpretation. Across all categories, 29.2\% of transitions following EWS were statistically significant at the $p < .05$ level, with an additional 37.8\% marginally significant at the $p < .1$ level. Notably, larger transitions showed the strongest evidence: for instance, nearly 59\% of large increases and 39\% of large decreases were significant, while over two-thirds of massive changes reached at least marginal significance. In contrast, small changes were rarely significant representing clear false negatives.

\section{Discussion}

CSD has been documented across diverse complex systems---including ecosystems, climate regimes, financial markets, and social networks---where early signs of instability precede abrupt shifts into alternative stable states \cite{scheffer2009early,scheffer2010complex}. In doing so, monitoring CSD has brought many advances to how we understand and forecast important phenomena long before they happen \cite{dakos2024tipping,fraccascia2018resilience,milkoreit2018defining,nazarimehr2020critical,olderikkert2016slowing}. To the best of our knowledge, this is the first study to report strong evidence of CSD in educational contexts. These findings open significant new possibilities for both research and practice. Below, we discuss the findings and their implications.
 
Our study found consistent evidence of critical slowing down in the vast majority of students (88.2\%), with a substantial number of detections per student prior to dropout. Warnings were rare in the initial phase but concentrated heavily in the final two quarters. This late-stage concentration was further corroborated by detailed timing analysis. The first strong signals typically appeared well past the midpoint of a student's engagement, with activity peaking late before tapering off as dropout approached. CSD consistently appeared as bursts rather than being distributed diffusely. Taken together, CSD metrics and their occurrence are converging lines of evidence that show that signals of critical slowing down are both evident, temporally concentrated towards the end of most students. What makes this evidence particularly compelling is that, regardless of the length of the time series, warning signals consistently appeared close to dropping out. Furthermore, while studies in other disciplines typically examine a single system or a few instances—such as stock prices or a handful of species, e.g., \cite{dakos2024tipping,milkoreit2018defining}, our study provided evidence in several thousand students.
\cite{dakos2024tipping,milkoreit2018defining,olderikkert2016slowing}.

Empirically, the consistency provides support that resilience loss in educational contexts follows the same universal warning dynamics observed in other complex systems, e.g., ecosystems, climate systems, and financial markets \cite{dakos2024tipping,milkoreit2018defining,olderikkert2016slowing, scheffer2009early, scheffer2020critical}. From a theoretical perspective, identifying CSD in education provides a novel framework for conceptualizing how learning dynamics evolve, destabilize, and potentially transition into qualitatively different states \citep{Saqr2026-ec}. In doing so, it advances our understanding of learning as a complex dynamic system, where periods of apparent stability may mask underlying vulnerabilities that can be detected through statistical indicators \cite{hilpert2018complex, scheffer2009early, scheffer2020critical}. From a practical standpoint, the early detection of resilience loss at the individual level creates opportunities for developing responsive interventions that can be deployed before learners reach critical points of collapse, disengagement, or derailment of other functions. 
Unlike many traditional learning analytics indicators, which often fail to generalize across different systems or datasets, the early warning signals identified here are grounded in fundamental resilience dynamics and apply widely \cite{dakos2024tipping, scheffer2009early, scheffer2020critical, olderikkert2016slowing}. The generic and universal nature of CSD indicators, coupled with their consistent emergence across a large number of students, suggests that they have the potential to be portable, offering a more generalizable tool for detecting loss of resilience and changes in diverse educational contexts. Possible applications include many digital and multimodal data sources that generate intensive time series data, such as physiological data. Signals such as heart rate variability, galvanic skin response, or EEG rhythms are already used to assess stress, attention, and cognitive load. 

In our study, 11.8\% of the participants had no detectable CSD warning signs before they were disengaged and dropped out. These can be collectively considered false negatives (or simply negatives). Several mechanisms can account for negative cases. First, a general limitation of EWS is that CSD only appears when changes in a system occur slowly relative to the system’s own timescale. For instance, if students shift abruptly following an external factor, a life event (moving out of school or town or a family event), disengagement occurs too rapidly for early warning signals to appear or be captured. Second, human systems are inherently more complex, stochastic, and less predictable compared to the physical or biological systems where CSD has been most extensively studied \cite{dakos2024tipping, swingedouw2020early, milkoreit2018defining,scheffer2020critical}. Furthermore, human systems involve agency. Unlike ecosystems or climate regimes, students can change their behavior in response to a multitude of factors. Furthermore, small perturbations (adversities) can also trigger disproportionate changes in nonlinear systems, making the transition inherently difficult to predict. Finally, it is important to note that early warning indicators do not capture every tipping point or every type of transition. These limitations highlight the need for complementary or alternative indicators to provide a fuller picture of resilience loss and impending critical transitions \cite{fraccascia2018resilience,scheffer2010complex,scheffer2009early,scheffer2020critical}.

False positives represent the opposite challenge, where apparent signals arise without an actual approach to a tipping point. Distinguishing true declines in resilience from heightened environmental noise is difficult, as both can generate changes in variance or autocorrelation that mimic CSD. It is also important to note that CSD signals reflect weakening stability in the current state, not the direction of the shift that follows  \cite{dakos2024tipping,scheffer2009early, EWSmethods, Dakos2015-zx}. For example, a student performing well at a certain proficiency level may begin to slow down as tasks become more difficult and eventually drop out. This could precede a drop in performance to a lower level, but also it could reflect the strain that comes before a breakthrough, where the student adapts and succeeds at a higher level. CSD as a mechanism does not distinguish between negative and positive transitions, it mainly captures impending major events. Interestingly, CSD has been mostly reported in cases where negative shifts occurred, e.g., collapse of a stock or an ecosystem which seems to be the case in our study too where most of the cases dropped out \cite{fraccascia2018resilience, dakos2024tipping}.

In resilience terms, a regime change is about a structural reorganization of the system where the system settles into a qualitatively different state \citep{fraccascia2018resilience}. In our sample, most of the supposedly false positives were in fact changes in performance levels. Almost 77.8\% of the students who had early EWS shifted to a new state where their grades underwent a major change. Regardless of where one chooses to draw the line, and even under more stringent assumptions, such as requiring statistical significance (which is not a formal prerequisite for defining a regime shift), the results hold. Many of these early warning signals still corresponded to genuine indicators of major shifts in math practice scores. 
It is very important to note here that we conducted the comparison (before vs. after signal value of practice scores) for the subgroup with early EWS, as these cases provided sufficient data to assess the nature of the subsequent transition. For the cohort exhibiting late EWS, the number of post-warning time points was insufficient to ensure adequate statistical power, as dropout occurred following EWS. This underscores the critical limitation of value-based approaches: by the time absolute performance metrics show decline, the transition may already be underway.

\subsection{Limitations}

Applying CSD to students’ learning brings several challenges. First, the data and observational requirements are demanding. Detecting changes in resilience with CSD requires a long, high-quality time series of learner behavior. Second, CSD's sensitivity to noise and non-stationarity complicates the picture. Furthermore, student learning behavior is inherently variable, influenced by context, peers, and external demands. This variability produces noise that can mimic the very patterns that CSD aims to capture. Third, while CSD is valued for its generic and universal character, this comes with the drawback of low specificity. CSD relies on the idea that systems approach tipping points slowly enough for resilience loss to accumulate. Human systems introduce unique complications. Unlike ecosystems or physical systems, students adapt to predictions and feedback. If early warning tools are made visible to learners or teachers, students may alter their behavior in reflexive ways, either masking the signal or creating new ones. In this sense, human agency and social dynamics can mute, distort, or even reverse CSD patterns, reducing predictability. Together, these limitations highlight both the promise and the caution required in applying critical slowing down to education. 
There is also a trade-off between specificity and sensitivity, lowering detection threshold would allow more positives cases on the expense of false positives.

\subsection{Conclusion}

This study provides the first robust empirical evidence of CSD in educational contexts, marking a significant advancement in both educational research and the broader application of complexity science. Our findings demonstrate that signals of resilience loss ---long recognized in ecosystems, climate systems, and financial markets--- also manifest in learning dynamics, particularly as students approach dropout. These signals were found in a vast majority of students, appearing consistently and with temporal concentration in the final stages of their engagement. The empirical consistency of these signals across thousands of students suggests that educational systems, like other complex systems, exhibit universal warning dynamics prior to transitions into alternative stable states. The detection of CSD highlights the potential for early warning systems based as resilience indicators in educational contexts.
Practically, the detection of CSD opens new opportunities for timely, personalized interventions aimed at supporting students before critical tipping points are reached. Moreover, the portability of CSD indicators across systems and data types raises the possibility of expanding this approach to include physiological and multimodal data sources, enabling real-time resilience monitoring and adaptive educational support.

\section*{Acknowledgment}

This work is co-funded by the Research Council of Finland (Suomen Akatemia) through the project TOPEILA, Decision Number 350560 which was received by the first author, and CRETIC, Decision Number 360746, which was received by the second author.

\bibliographystyle{abbrv}
\bibliography{references}  






\end{document}